\documentclass[slac_one]{revtex4}

\usepackage{graphicx}

\usepackage{fancyhdr}

\newcommand{\ga}{\alpha}

\newcommand{\gs}{\sigma}

\newcommand{\gD}{\Delta}

\begin{document}


\title{{\small{2005 International Linear Collider Workshop - Stanford, U.S.A.}}\\ 

\vspace{12pt}

Experimental studies of Strong Electroweak Symmetry Breaking in gauge boson scattering and three gauge boson production} 


%




\author{P. Krstonosic}

\affiliation{DESY-Hamburg, Hamburg, 22607, Germany}

\author{K. M\" onig}

\affiliation{LAL-Orsay and DESY-Zeuthen, Zeuthen, 15738, Germany}

%

\author{M. Beyer}

\affiliation{Institute of Physics, University of Rostock, 18051 Rostock,
  Germany}

\author{E. Schmidt}
\affiliation{Institute of Physics, University of Rostock, 18051 Rostock,
  Germany}

\author{H. Schr\" oder}
\affiliation{Institute of Physics, University of Rostock, 18051 Rostock,
  Germany}

\begin{abstract}

Abstract. If no light Higgs boson exist, the interaction among the gauge bosons
becomes strong at high energies ($\sim$ 1TeV). The effects of strong
electroweak symmetry breaking (SEWSB) could manifest themselves as anomalous
couplings before they give rise to new physical states, thus measurement of
all couplings and their possible deviation from Standard Model (SM) values
could give valuable information for understanding the true nature of symmetry
breaking sector. Here we present a detail study of the measurement of quartic
gauge couplings in weak boson scattering processes and a possibility for 
same measurement in triple weak boson production. Expected limits on the parameters ${{\alpha}_4}, {{\alpha}_5},{{\alpha}_6},{{\alpha}_7}$ and ${{\alpha}_{10}} $ in electroweak chiral Lagrangian are given.

\end{abstract}


\maketitle

\thispagestyle{fancy}





\section{INTRODUCTION} 

\begin{table}[b]

\begin{center}

\caption{Sensitivity to quartic anomalous couplings for all possible scattering processes}

\begin{tabular}{|l|l|c|c|c|c|c|}

\hline \textbf{$ {e^+}{e^-}\rightarrow $} & \textbf{ $ {e^-}{e^-}\rightarrow $} & \textbf{$ {\alpha}_4 $}
 & \textbf{${\alpha}_5 $}  & \textbf{${\alpha}_6 $}  & \textbf{$ {\alpha}_7 $}  & \textbf{${\alpha}_{10} $}
\\

\hline ${W^+}{W^-}\rightarrow {W^+}{W^-} $ & $ {W^-}{W^-}\rightarrow {W^-}{W^-} $ & + & + &  &  &  \\

\hline ${W^+}{W^-}\rightarrow ZZ $ &  & + & + & + & +  &  \\

\hline ${W^{\pm}}Z\rightarrow {W^{\pm}}Z $ & ${W^-}Z\rightarrow {W^-}Z $ & + & + & + & +  &  \\

\hline $ ZZ\rightarrow ZZ $ & $ ZZ\rightarrow ZZ $ & + & + & + & +  & + \\

\hline

\end{tabular}

\label{uzas}

\end{center}

\end{table}

Investigation of the true mechanism that triggers electroweak symmetry breaking is one of the most intriguing questions that are still open in the SM. If we start with minimal set of assumptions and adopt an effective Lagrangian approach~\cite{bot_up} then all our knowledge and ignorance about the underlying working theory is parameterized in set of couplings that are associated with operators making the theory finite to a given order. In the Higgs-less scenario anomalous quartic couplings  ${{\alpha}_4}, {{\alpha}_5},{{\alpha}_6},{{\alpha}_7}$ and ${{\alpha}_{10}} $ of (longitudinally polarized) W and Z boson are of interest since they give information on the underlying dynamic of the new symmetry. The set of operators that we consider 
\begin{equation}\label{eq:L4}
\hbox{$L_4={{\alpha}_4 \over 16{\pi}^2} tr(V_{\mu}V_{\nu})tr(V^{\mu}V^{\nu})$ \ ,\ 
$L_5={{\alpha}_5 \over 16{\pi}^2}  tr(V_{\mu}V^{\mu})tr(V_{\nu}V^{\nu})$}  
\end{equation}
\begin{equation}\label{eq:L5}
\hbox{$L_6={{\alpha}_6 \over 16{\pi}^2}tr(V_{\mu}V_{\nu})tr(TV_{\mu}))tr(TV^{\nu})$ \ ,\ 
$L_7={{\alpha}_7 \over 16{\pi}^2}tr(V_{\mu}V_{\mu})tr(TV_{\nu}))tr(TV^{\nu})$ \ ,\ 
$L_{10}={{\alpha}_{10} \over 32{\pi}^2}(tr(TV_{\mu}))tr(TV_{\nu}))^2$  }  
\end{equation} 
uses the same notation as in ~\cite{kjer}, thus the results can be directly compared. Equation (~\ref{eq:L4}) contains  SU(2)$\sb{c}$ conserving operators and (~\ref{eq:L5}) ones that are allowed if isospin symmetry is not conserved. Full set of possible scattering processes that could be used to extract the couplings is given in Table~\ref{uzas}, where we have tried to cover all of them with at least one decay channel (fully hadronic). Only in the multi-parameter analysis do mutual interplay of the couplings come in to the result, thus making the prediction more realistic.

\section{VECTOR BOSON SCATTERING}

\subsection{General Layout}

We assume a  center of mass energy of 1TeV and a  total luminosity of
1000fb$\sp{-1}$ in $ {e^+}{e^-}$ and 350fb$\sp{-1}$ in ${e^-}{e^-}$ mode. Beam
polarization of 80\% for electrons and 40\% for positrons is also assumed. The six
fermion processes under study correspond to the scattering of longitudinal
weak bosonos. Since triple weak boson production is also sensitive to quartic
anomalous couplings ( $ZZ$ or ${W^+}{W^-}$ with neutrinos of second and third
generation as well as a part of ${{\nu}_{e}}\bar {\nu}_{e}WW(ZZ)$,
$e{\nu}_{e}WZ$ and ${e^+}{e^-}{W^+}{W^-}$  final states ) there is no distinct
separation of signal and background. Signal processes in separate analysis are
thus affected by all other signal processes as well as pure background. In
comparison to the previous study~\cite{kjer} single weak boson production was
included in background for completeness and in order to get closer to
the experimental conditions initial state radiation was taken into account when
generating events. For the generation of  $t\bar t$ events
Pythia~\cite{pythia} was used, for all other processes the full six fermion
generator WHIZARD~\cite{whizard} was used. No flavor summation was done since
all possible quark final states were generated. Hadronisation was done with
Pythia. The SIMDET~\cite{simdet} program was used to producethe detector response of a
possible ILC detector. Table~\ref{l2ea4-t2} contains a summary of all generated
processes used for analysis and corresponding cross sections. For pure
background processes a full 1ab$\sp{-1}$ sample was generated, all signal
processes were generated with higher statistics. Single weak boson processes
and $q{\bar q}$ events were generated with an additional cut on
M($q{\bar q}$)$>$130GeV to reduce number of generated events.

\begin{table}[t]

\begin{center}

\caption{Generated processes and cross sections of signal and background for $\sqrt{s}=1$TeV, polarization 80\% left for electron and 40\% right for positron beam }

\begin{tabular}{|l|r|l|r|}

\hline \textbf{Channel} & \textbf{ $\sigma [fb] $ \ } &  \textbf{Channel} & \textbf{ $\sigma [fb] $ \ }
\\

\hline \  ${e^+}{e^-}\rightarrow{{\nu}_{e}} {{\bar {\nu}}_{e}}{W^+}{W^-}\rightarrow{{\nu}_{e}}{\bar {\nu}}_{e}q{\bar q}q{\bar q}$ & \ 23.19 & \ ${e^-}{e^-}\rightarrow{{\nu}_{e}} {{\bar {\nu}}_{e}}{W^-}{W^-}\rightarrow{{\nu}_{e}}{\bar {\nu}}_{e}q{\bar q}q{\bar q}$ & 27.964  \\

\hline \ ${e^+}{e^-}\rightarrow{{\nu}_{e}} {{\bar {\nu}}_{e}}ZZ\rightarrow{\nu}_{e}
{\bar {{\nu}_{e}}}q{\bar q}q{\bar q}$ & 7.624 & \  ${e^-}{e^-}\rightarrow{e^-}{{\nu}_e}{W^-}Z\rightarrow{e^-}{{\nu}_e} q{\bar q}q{\bar q}$ & 80.2  \\

\hline \ ${e^+}{e^-}\rightarrow{\nu}{\bar {\nu}}q{\bar q}q{\bar q}$ (3V contribution) & 9.344 & \  ${e^-}{e^-}\rightarrow{e^-}{e^-}ZZ\rightarrow{e^-}{e^-} q{\bar q}q{\bar q}$ & 3.16 \\

\hline \ ${e^+}{e^-}\rightarrow{{\nu}}e{W}Z\rightarrow{{\nu}}e q{\bar q}q{\bar q}$ & 132.3 & \  ${e^-}{e^-}\rightarrow{e^-}{e^-}{W^+}{W^-}\rightarrow{e^-}{e^-} q{\bar q}q{\bar q}$ & 443.9  \\

\hline \ ${e^+}{e^-}\rightarrow{e^+}{e^-}ZZ\rightarrow{e^+}{e^-} q{\bar q}q{\bar q}$ & 2.09 & \ $ {e^-}{e^-}\rightarrow{e^-}{e^-}t{\bar t} \rightarrow{e^-}{e^-} X$ & 0.774   \\

\hline \ ${e^+}{e^-}\rightarrow{e^+}{e^-}{W^+}{W^-}\rightarrow{e^+}{e^-} q{\bar q}q{\bar q}$ & 414.6 & \  ${e^-}{e^-}\rightarrow ZZ \rightarrow q{\bar q}q{\bar q}$ & 232.875  \\

\hline \  ${e^+}{e^-}\rightarrow t{\bar t} \rightarrow X$ & \ 331.768  & \ ${e^-}{e^-}\rightarrow {e^-}{{\nu}_e}{W^-} \rightarrow {e^-}{{\nu}_e}q{\bar q}$ & \ 235.283  \\

\hline \ ${e^+}{e^-}\rightarrow {W^+}{W^-} \rightarrow q{\bar q}q{\bar q}$ & \ 3560.108 & \  ${e^-}{e^-}\rightarrow {e^-}{e^-}Z \rightarrow {e^-}{e^-}q{\bar q}$ & 125.59 \\

\hline \ ${e^+}{e^-}\rightarrow ZZ \rightarrow q{\bar q}q{\bar q}$ & 173.221& & \\

\hline \ ${e^+}{e^-}\rightarrow e{\nu}W \rightarrow e{\nu}q{\bar q}$ & 279.588 & & \\

\hline \ ${e^+}{e^-}\rightarrow {e^+}{e^-}Z \rightarrow {e^+}{e^-}q{\bar q}$ & 134.935 & &  \\

\hline \ ${e^+}{e^-}\rightarrow q{\bar q}\rightarrow X$ & 1637.405 & &  \\

\hline

\end{tabular}

\label{l2ea4-t2}

\end{center}

\end{table}


The observables sensitive to the quartic couplings are the total cross section (either reduction or increase depending on the interference term in the amplitude and the point in parameter space), and modification of the differential corross section distributions over polar angle as well over decay angle. This is not a full set of observables but some sensitive event variables, for example transverse momentum, cannot be used since contribution of longitudinally polarized weak bosons is dropping faster then for transversally polarized wak bosons with increasing transverse momentum and a transverse momentum cut is an unavoidable tool to suppress background in analysis.

\subsection{Event selection}

\begin{figure}[t]
\begin{minipage}[h]{70mm}
\includegraphics[width=55mm,bb=106 251 497 599]{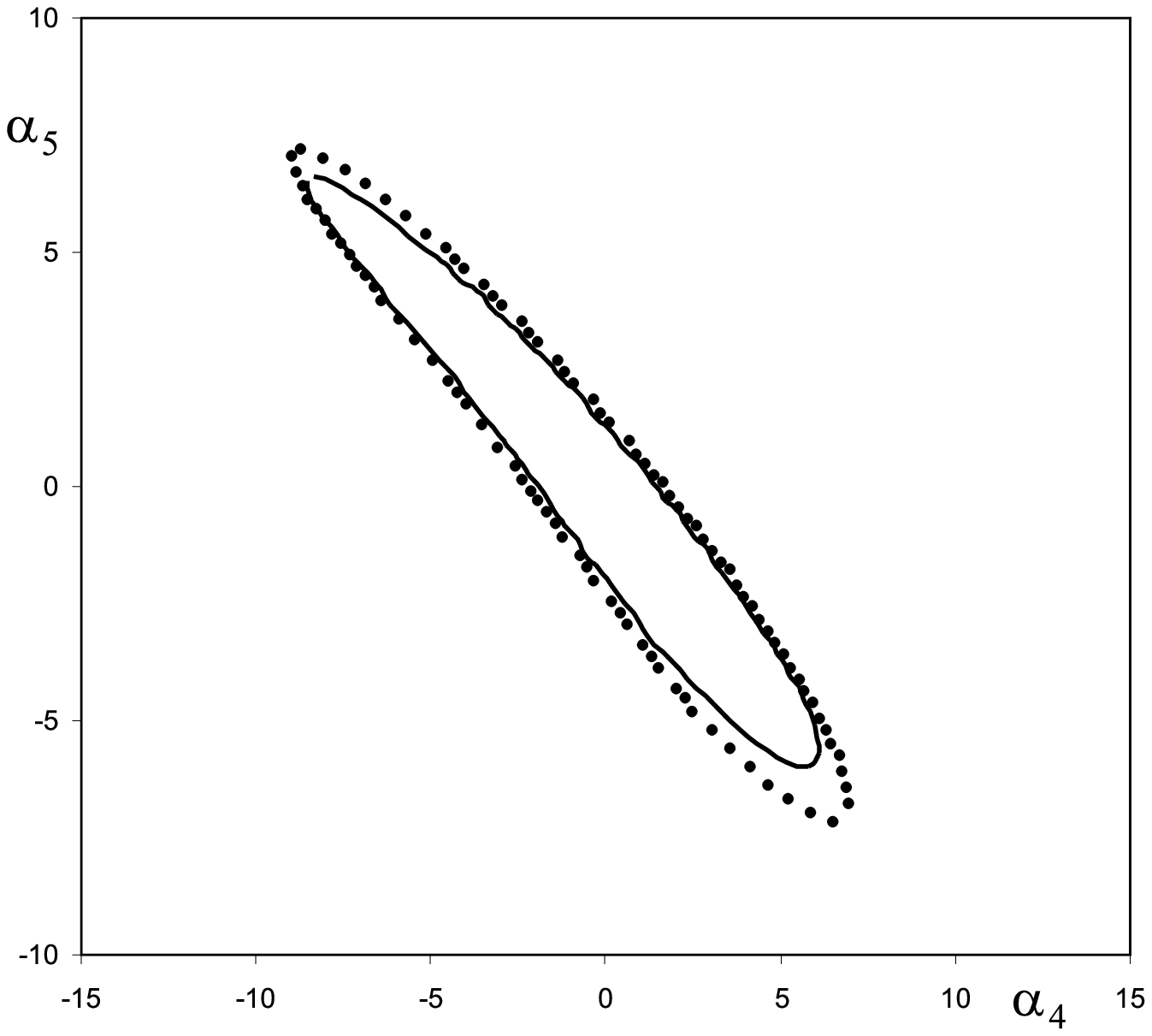}
\caption{The influence of ISR on result for $W^+W^-$ scattering at 800GeV, 68\% confidence level contour, full line with ISR, dotted no ISR}
\label{fig:poredjenje}
\end{minipage}
\hspace{5mm}
\begin{minipage}[h]{90mm}
\includegraphics[width=51mm,bb=113 240 477 591]{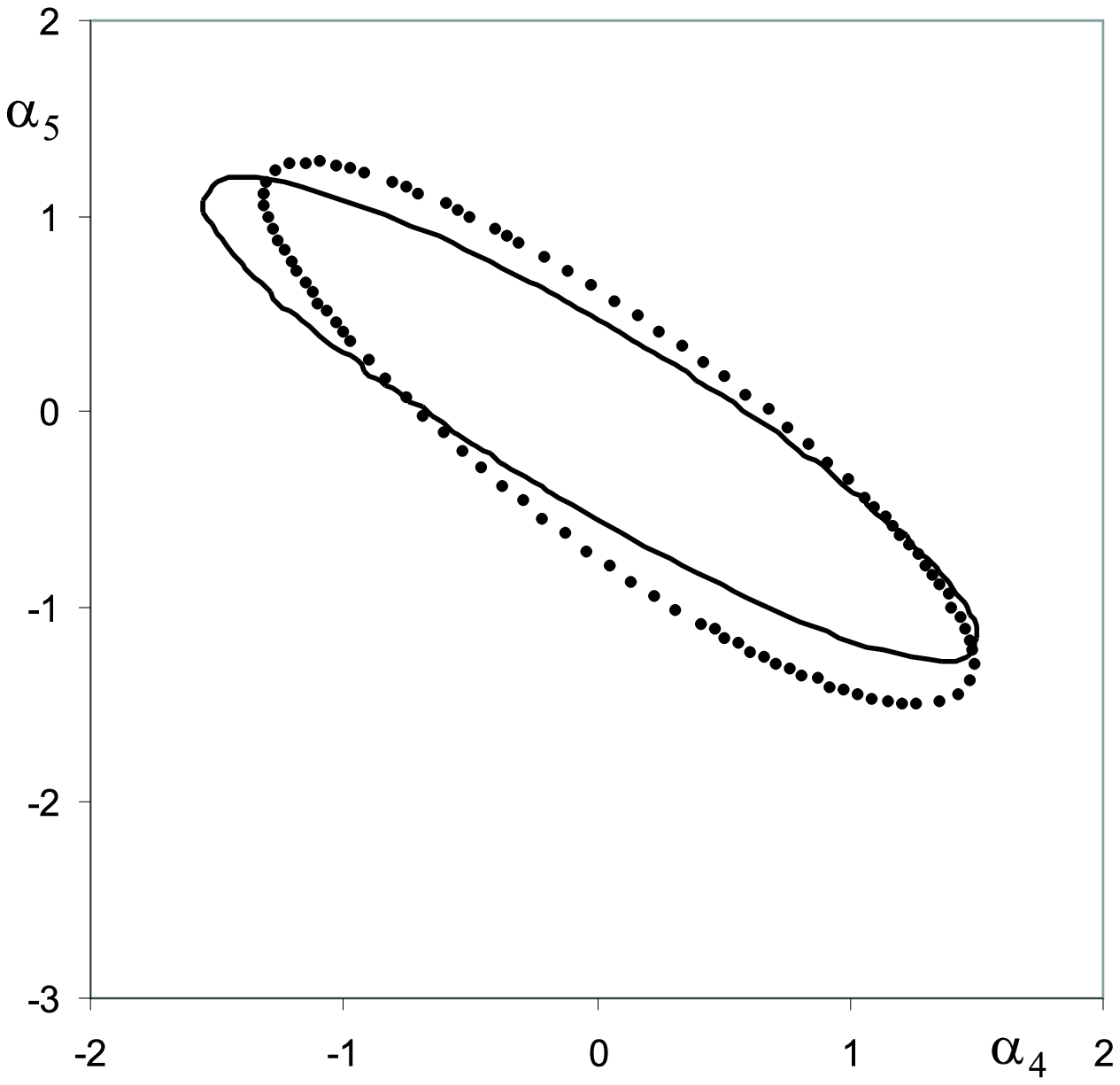}
\caption{Reachable sensitivity using only the $e^+e^-$ mode (full line) and combination of $e^+e^-$ and  $e^-e^-$ equally sharing time (dotted line) at 1TeV, 68\% confidence level contour}
\label{fig:poredjenje2}
\end{minipage}
\end{figure}

Event selection was done using a cut based approach similar to previous analysis ~\cite{kjer}. The general steps in the analysis were the  use of final state $e^-$($e^+$) to tag background (signal in $e{\nu}_{e}WZ$ case), a  cut on transverse momentum, and missing mass and energy. Realistic ZVTOP b-tagging ~\cite{ztop} was used when possible to enhance signal to background separation. Finally cuts around the nominal masses of weak boson were used to accept only well reconstructed events.
The effects of the introduction of ISR were investigated at center of mass energy of
800GeV and $W^+W^-$ scattering as an example. The signal event sample was generated with and without ISR and the fit results were compared without taking background into
consideration. The difference in the fit result, Fig.~\ref{fig:poredjenje} show, as
expected, deterioration of the result to a small extent. The significant effect of ISR
is on the smearing of kinematic distributions and making harder signal
background separation.
\subsection{Fit method and results}

Extraction of quartic gauge couplings from reconstructed kinematic variables was done with a binned likelihood fit. For each signal process statistics much larger than the nominal one (1000fb$\sp{-1}$ for ${e^+}{e^-}$) were 
generated and passed through the detector simulation. Each event is described by reconstructing four kinematic 
variables - event mass, absolute value of production angle cosine and absolute values of decay angle cosines  
of each reconstructed weak boson. The absolute value of the production and decay angles were used since there is no possibility to resolve quark antiquark and ${W^+}{W^-}$ ambiguities. Matrix element calculation from WHIZARD was used to obtain weights to reweight the event as a function of quartic gauge couplings. Each Monte Carlo 
SM event is weighted by:  
\begin{equation}\label{eq:units1}
\hbox{$R({\alpha}_i,{\alpha}_j)=1+A{{\alpha}_i}+B{{\alpha}_i}^2+C{{\alpha}_j}+D{{\alpha}_j}^2+E{{\alpha}_i}{{\alpha}_j}$}  
\end{equation}
Function $R({\alpha}_i,{\alpha}_j)$ describes the quadratic dependence of the differential cross-section on the
 couplings. It is obtained in the following way: using the generated SM events (${\alpha}_i=0 , i=4,5,6,7,10$) 
we recalculated the matrix elements of the event for a set of five different points in ${\alpha}_i,{\alpha}_j$
 space and solve a set of linear equations for A,B,C,D and E. Due to the linear combination in which couplings 
contribute in amplitude ~\cite{helamp}, in any case five points are enough to determine the constants for weighting function. Choice of the points varied from process to process in order to fulfill the following conditions: 
distance of the point from SM value should be large enough not to come in numerical instabilities problem 
when solving equations and at the same time small enough not to come in to the region were phase space population would be significantly different from the SM. Four dimensional event distributions are fitted with 
MINUIT~\cite{minuit} maximizing the likelihood as a function of  ${\alpha}_i$,${\alpha}_j$ taking 
the SM Monte Carlo sample as ``data''.
\begin{equation}\label{eq:units2}
\hbox{$L(\alpha_p,\alpha_q)=-\sum\limits_{i,j,k,l}{N^{SM}}(i,j,k,l) \ln\left( {N^{{\alpha}_p,{\alpha}_q}{}}(i,j,k,l)\right )+\sum\limits_{i,j,k,l} {N^{{\alpha}_p,{\alpha}_q}}(i,j,k,l)$}  
\end{equation}
where $i$ runs over the reconstructed event energy, $j$ over the production angle, $k$
and $l$ over the decay angles,${N^{SM}}(i,j,k,l)$ are the ``data'' which
coresspond to the SM Monte Carlo sample and $
{N^{{\alpha}_p,{\alpha}_q}}(i,j,k,l) $ is the sum of same SM events in the bin
each weighted by $R({\alpha}_p,{\alpha}_q)$. Pure background events have
$R({\alpha}_p,{\alpha}_q)=1$, and for background coming from other sensitive
processes the proper weight is taken into account. After separate analysis of each
process from Table~\ref{uzas} a combined fit was done. Small fraction of double
counted events that remains after single process analysis was uniquely
assigned to one or another set according to the distance from the nominal mass of
the weak boson pair ( for example WW or ZZ ).
The analysis was primarily focused on $e^+e^-$  mode since we expect
significantly larger integrated luminosity in this mode. Effects of possible
$e^-e^-$ option were considered in following way. If available time for running
the collider at a given center of mass energy is fixed and $e^-e^-$ option exists
we can divide running time between $e^+e^-$ and $e^-e^-$ assuming that the ratio of
their integrated luminosities is 3:1 and then do the fit to the whole data
sample. A combined fit for 2ab$\sp-1$ $e^+e^-$ was done and compared with
the combined result from 1ab$\sp-1$ $e^+e^-$ together with a 350fb$\sp-1$ $e^-e^-$
sample. The confidence level contours in Fig.~\ref{fig:poredjenje2} show negligible
difference in reachable sensitivity in these two cases.

\begin{figure*}[t]
\centering
\includegraphics[width=55mm,bb=131 323 452 620]{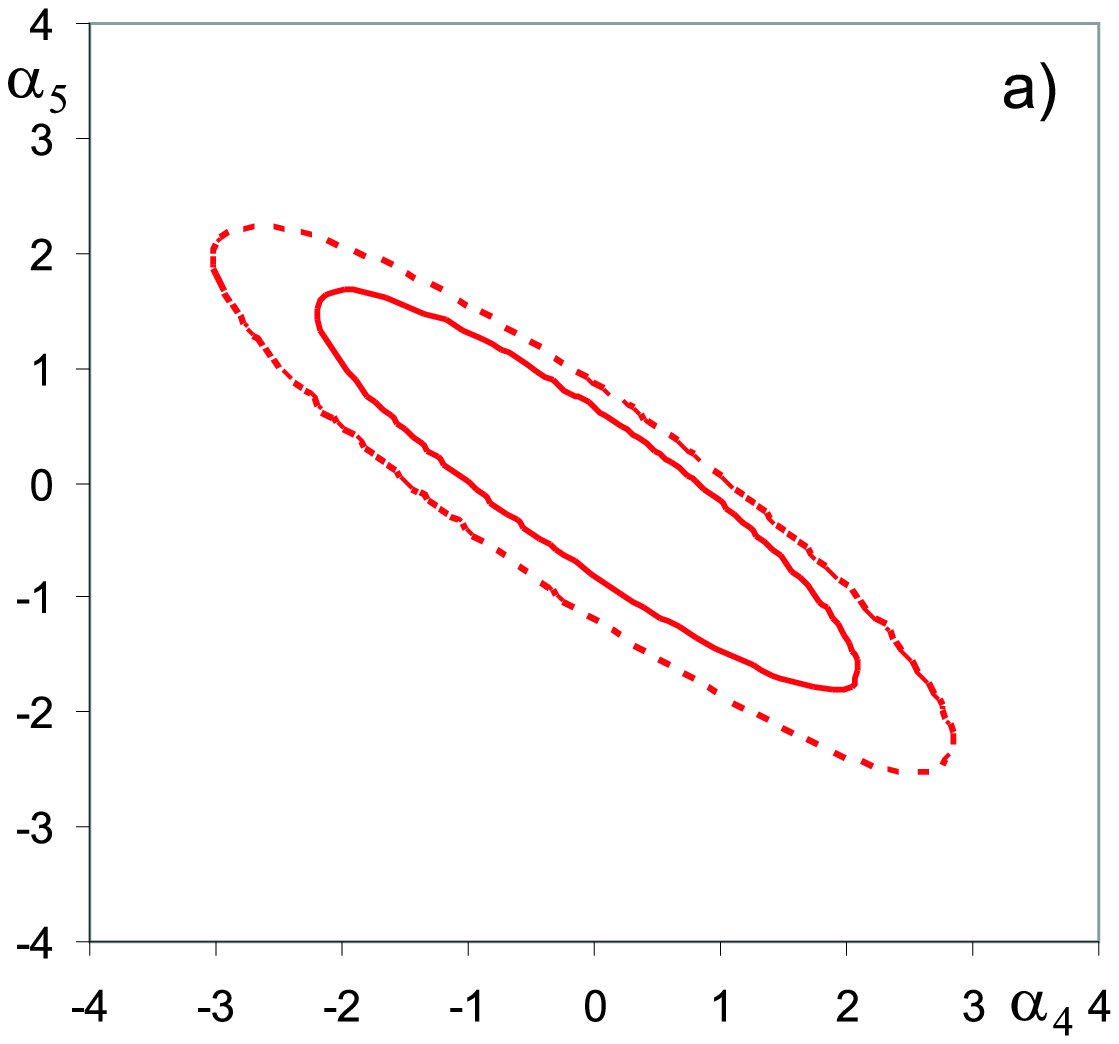}
\includegraphics[width=110mm,bb=128 329 808 627]{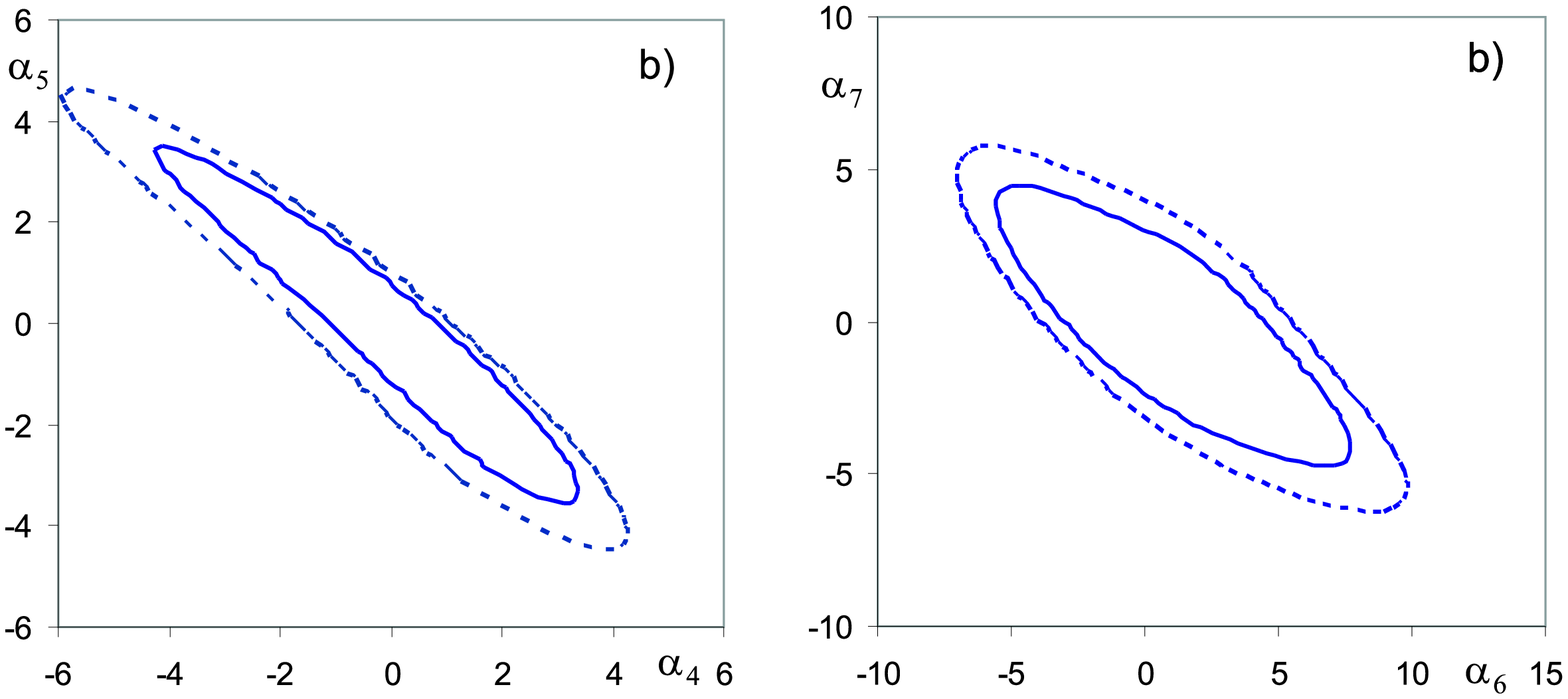}
\caption{Expected sensitivity (combined fit to all sensitive processes) to quartic anomalous couplings form 1000fb$\sp{-1}$ $e^+e^-$ sample  full line (inner one) represents 68\% and dotted (outer one) 90\% confidence level  a) conserved  SU(2)$\sb{c}$ case  b) broken   SU(2)$\sb{c}$ case  }
\label{fig:kskksk}
\end{figure*}
%
%
\begin{table}[h]
\begin{minipage}[t]{0.45\textwidth}
\caption{The expected sensitivity from 1000fb$\sp{-1} e^+e^-$ sample at 1TeV in
  SU(2)$\sb{c}$ conserving case, positive and negative one sigma errors given separately.}
\begin{tabular}{|c|c|c|}

\hline \textbf{\ coupling \ } & \textbf{ $\sigma -$} & \textbf{$\sigma + $}\\

\hline \ $\alpha_4$ \ & \ -1.41 \  & \ 1.38 \ \\

\hline \ $\alpha_5$ \ & \ -1.16 \  & \ 1.09 \  \\

\hline

\end{tabular}
\label{trtmrt}

\end{minipage}
%
\hspace{.25cm}
\begin{minipage}[t]{0.45\textwidth}
\caption{The expected sensitivity from 1000fb$\sp{-1} e^+e^-$ sample at 1TeV in
  broken SU(2)$\sb{c}$ case, positive and negative one sigma errors given separately.}
\begin{tabular}{|c|c|c|}

\hline \textbf{\ coupling \ } & \textbf{ $\sigma -$} & \textbf{$\sigma + $}\\

\hline \ $\alpha_4$ \ & \ -2.72 \  & \ 2.37 \ \\

\hline \ $\alpha_5$ \ & \ -2.46 \  & \ 2.35 \  \\

\hline \ $\alpha_6$ \ & \ -3.93 \  & \ 5.53 \ \\

\hline \ $\alpha_7$ \ & \ -3.22 \  & \ 3.31 \  \\

\hline \ $\alpha_{10}$ \ & \ -5.55 \  & \ 4.55 \ \\

\hline

\end{tabular}
\label{krk-t4}

\end{minipage}
\end{table}

%
Table~\ref{trtmrt} and Table~\ref{krk-t4} contain results for weak
boson scattering assuming  integrated luminosity of 1000fb$\sp{-1}$ in  ${e^+}{e^-}$ mode in SU(2)$\sb{c}$ conserving case and broken  SU(2)$\sb{c}$.

\section{TRIPLE BOSON PRODUCTION}

We consider now the reactions $e^+e^-\rightarrow W^+W^-Z$ and
$e^+e^-\rightarrow ZZZ$. On tree level the elementary process producing the
$WWZ$ final state is driven by 15 Feynman diagrams.  Only one of
the diagrams contains the quartic coupling and has to be extracted from the
other interfering terms. Only the part containing a longitudinal gauge boson is
expected to give a sizable signal related to electroweak symmetry breaking.
For $WWZ$ this part is substantially enhanced using polarized beams. We investigate
several cases: i) unpolarized, ii) $80\%$ right polarized electrons, and iii)
$80\%$ right polarized electrons along with $60\%$ left polarized positrons.
For $ZZZ$ polarization is not substantial, since the standard model background
is much smaller (two diagrams). Calculations are done using the Whizard event
generator~\cite{whizard}.  Since the gauge bosons are short lived
states they decay.  Presently we consider on-shell gauge bosons only (narrow
width approximation) and hadronize the final state using
PYTHIA~\cite{pythia}.The three-boson state is characterized by three four-momenta and the bosonic spins. In
general three momenta lead to 12 kinematical variables that are reduced by
four through energy momentum conservation, by three because of the on-shell
condition mentioned before, and by two due to rotational invariance. Hence in
total three independent kinematical variables are left. We choose two
invariant masses of the Dalitz plot, $M_{WZ}^2$, $M_{WW}^2$ and the angle
$\theta$ between the beam axis and the direction of the $Z$-boson. Spin of the
bosons leads to additional degrees of freedom, and we may distinguish
longitudinal ($L$) from transverse ($T$) polarization.
Presently, we do not yet analyze the bosonic spins.
\begin{figure*}[b]
\centering
 
 \includegraphics[width=55mm,bb=70 571 230 674]{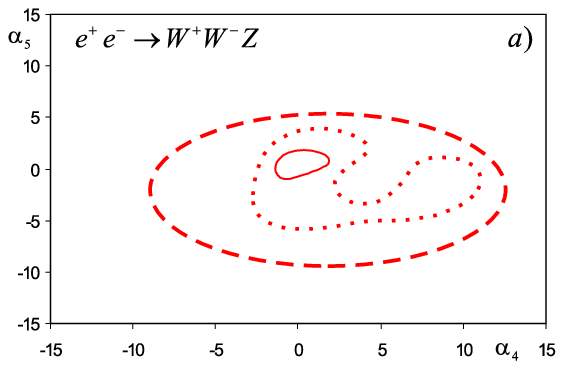}
 \includegraphics[width=55mm,bb=70 571 230 674]{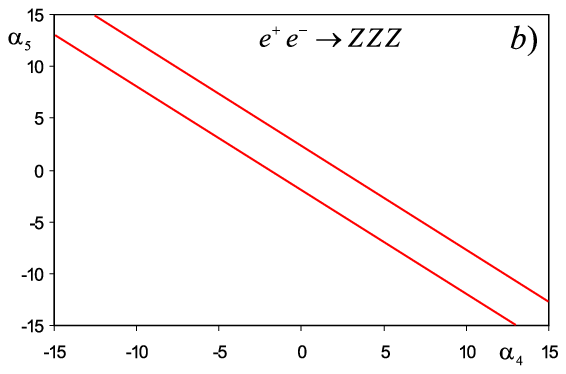}
\caption{Expected sensitivity for $\ga_4$ and $\ga_5$ at $\sqrt{s}=1000$ GeV.
  The lines represent $90\%$ confidence level. Luminosity assumption 1000
  fb$^{-1}$.  a) $e^+e^-\rightarrow WWZ$: 
  unpolarized case dashed line , $e^-$ right-polarized to $80\%$ dotted line , $e^+$
  additionally left-polarized to $60\%$ full line b) $e^+e^-\rightarrow ZZZ$
  unpolarized.} \label{fig:result}
\end{figure*}

The three independent kinematical variables lead to a three
dimensional histogram. If the angle $\theta$ is not measured the
resulting two dimensional histogram leads a Dalitz plot. We
investigate the differences on the histograms as a function of $\ga_4$
and $\ga_5$. The observable are discretize into bins and $\chi^2$ is given by
\begin{equation}
\chi^2=\sum_{i,j,k}\frac{N^{\rm exp}_{ijk}-N^{\rm theo}_{ijk}(\ga_4,\ga_5)}
{\gs_{ijk}^2}
\end{equation}
where $\gs_{ijk}$ denotes the error, and $i,j,k$ the sums over bins of
$M_{WZ}^2$, $M_{WW}^2$, and $\theta$. 

We use the Whizard generator to produce standard model events corresponding to
a luminosity of 1000 fb$^{-1}$.  The detector efficiency is simulated using
the fast simulation SIMDET~\cite{simdet}. To reconstruct $WWZ$, we use
the decay  $WWZ\rightarrow 6$ jets. The dominant background is due to $t\bar
t\rightarrow b\bar b WW\rightarrow 6$ jets. In the absence of a full simulation at
1000 GeV we estimate from our previous studies at 500 GeV~\cite{Beyer:2004km}
a combined effect of efficiency and purity to be 42\%. A full simulation at
1000 GeV is presently under development. The probabilities of standard model events are
reweighted when introducing anomalous couplings $\ga_4$, $\ga_5$. Since the
effective Lagrangian is linear in $\ga_4$, $\ga_5$ any observable is of second
order in the parameters and can be expressed by a polynomial with five
parameters. The parameters are determined by evaluation of $N^{\rm
  theo}_{ijk}(\ga_4,\ga_5)$ for each event and for five pairs of fixed values
$(\ga_4,\ga_5)$. By inversion $N^{\rm theo}_{ijk}(\ga_4,\ga_5)$ is know for
arbitrary values of $(\ga_4,\ga_5)$, viz.
\begin{equation}
N^{\rm  theo}_{ijk}(\ga_4,\ga_5)=
N_{ijk}^{\rm sm}+N_{ijk}^{\rm A}\ga_4+
N_{ijk}^{\rm B}\ga_4^2+N_{ijk}^{\rm C}\ga_5+
N_{ijk}^{\rm D}\ga_5^2+N_{ijk}^{\rm E}\ga_4\ga_5
\end{equation}
for each bin $i,j,k$. Finally we calculate
$\chi^2$ and determine $\gD\ga_4(\ga_4,\ga_5)$ and
$\gD\ga_5(\ga_4,\ga_5)$ for the specific values $\chi^2=2.30$ (68.3\%
confidence) and $\chi^2=4.61$ (90\% confidence). 
Results are shown in Fig.~\ref{fig:result}. We find that the sensitivity
drastically increases with polarization. Sensitivity can be improved even
further by utilizing meaningful cuts, which has not been done in the present
stage of analysis and by using the information of angular distribution of the jets that depends on the polarization stage of the final bosons.

\section{Conclusion}

\begin{figure*}[t]
\centering 
 \includegraphics[width=75mm,bb=52 363 455 731]{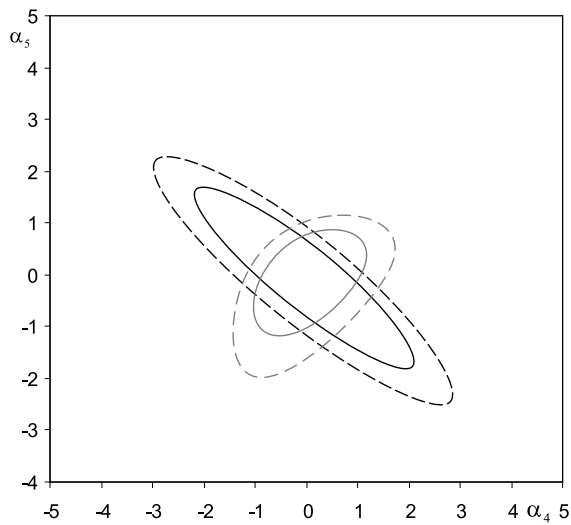}
\caption{Comparison of estimated sensitivities for $\ga_4$ and $\ga_5$ at
  $\sqrt{s}=1000$GeV from weak boson scattering (black) ad triple boson
  production (gray). Lines represent $68\%$ (full) and $90\%$ (deshed)
  confidence level contours.}
\label{fig:upsala}
\end{figure*}
Expected limits on the measurement of anomalous quartic couplings from all possible
 weak boson scattering processes were presented. In Fig.~\ref{fig:upsala} we make a
 comparison of estimated sensitivites from weak boson scattering processes and
 ongoing triple boson production analysis. With the same integrated luminosity
 and $80\%$ left $e^-$ and $40\%$ right $e^+$ polarization for scattering and 
 $80\%$ right $e^+$ and $60\%$ left $e^-$ polarization for triple producstion
 we obtain comparable results. This shows that luminosity sharing of opposite
 polarization can probably lead to the same overall accuracy for the measurement
 of quartic boson couplings. The same luminosity based conclusion was made
 after comparison of $e^+e^-$ and $e^-e^-$ running modes leaving the
 experimental physicist several ways to achieve the desired precision.

\end{document}